\documentclass[amsmath,amssymb,twocolumn]{revtex4}
\usepackage{dcolumn}
\usepackage{bm}
\usepackage{amsmath,mathrsfs}
\usepackage{graphicx,epsfig}
\usepackage{hyperref}
\usepackage{epsf}
\usepackage{color}

\begin{document}


\title{{\bf Wormholes in $f(R,T)$ gravity with variable equation of state}}

\author{S. Rastgoo}\email{rastgoo@sirjantech.ac.ir}
\author{ F. Parsaei }\email{fparsaei@gmail.com}

\affiliation{ Physics Department , Sirjan University of Technology, Sirjan 78137, Iran.}

\date{\today}


\begin{abstract}
\par    In this work, we introduce a novel set of asymptotically flat wormhole solutions within the framework of $f(R,T)$ theory of gravity. Considering a linear
 $f(R,T)=R+ 2\lambda T$ form, we show that a wide variety of wormhole solutions with asymptotically linear equation of state exist. Our solutions  satisfy all the energy conditions, namely the null, weak, strong and dominant energy conditions. The relationship between free parameters in the shape function and boundary conditions is analyzed. \\
\end{abstract}

\maketitle
\section{Introduction}
Theoretically, wormholes are solutions of the Einstein’s field equations that can connect two different points in the Universe. They can also serve as shortcuts between points in different universes \cite{Visser}. Flamm initially proposed the concept of a wormhole \cite{flamm}. Einstein and Rosen described the
structure of the Einstein-Rosen bridge mathematically, but the Einstein-Rosen bridge is not a traversable wormhole \cite{Rosen}. The term “wormhole” was introduced by Misner and Wheeler in 1957 \cite{wheeler}. In 1988, Morris and Thorne demonstrated that wormholes could be large enough for humanoid travelers and even allow for time travel  \cite{WH}. A traversable wormhole solution should not contain any horizon or singularity \cite{Visser}.
An exotic fluid that violates the null energy condition (NEC) in general relativity (GR) must be used to develop wormhole structures. This is the main ingredient of the wormhole theory in GR.
Numerous astronomical probes have confirmed the phenomenon of accelerating cosmic expansion, which has become a focus of interest for researchers in modern cosmology. Phantom wormholes violate the energy conditions (ECs) but have been studied extensively in the literature after the discovery of the accelerated expansion of the Universe\cite{phantom, phantom2, phantom1}. Many authors try to minimize the usage of exotic matter in constructing a wormhole. Thin shell wormhole \cite{cut, cut1, cut2}, wormhole with variable equation of state (EoS) \cite{Remo, variable}, and wormhole with polynomial EoS \cite{foad} are some examples. In all of these models, exotic matter is limited to a small region of spacetime.

 Modified theories of gravity are an alternative to describe the new cosmological problems. The additional levels of freedom, presented in these theories, can be reinterpreted as a generalized geometrical fluid.  This fluid holds distinct interpretations compared to the conventional matter fluids typically used as inputs in the field equations. A strong alternative to solving the problem of exotic matter can be found in modified theories. In this realm, wormholes have been studied extensively in modified theories. Wormholes are investigated in  Braneworld \cite{b, b1, b2, b3}, Born-Infeld theory \cite{Bo, Bo1}, quadratic gravity \cite{quad, quad1}, Einstein-Cartan gravity \cite{Cartan, Cartan1},  $f(Q)$ gravity \cite{fq, fq1, fq2, fq3, fq4} , $f(R)$ gravity \cite{fR0, fR11, fR22, fR33, fR44}.

In $f(R)$ gravity, the standard Einstein-Hilbert action is replaced by an arbitrary function of the Ricci scalar ($R$) \cite{fRR}. Additionally, $f (R, T)$ theory of gravity is generated by coupling any function of the Ricci scalar $R$ with the Lagrangian density of matter $Lm$ \cite{RT}. The $f(R,T)$ gravity can be seen as an extension of the  $f(R)$ modified theory of gravity \cite{RT}. The cosmological aspects within the framework of $f(R, T)$ have been explored in the literature \cite{cosm, cosm1, cosm2}. Black holes have been studied in the context of $f(R, T)$ gravity \cite{Black}.
 Various investigations have been carried out on wormhole solutions within $f(R,T)$ gravity. Azizi \cite{Azizi} has determined a shape function under the assumption of a linear EoS for matter. Solutions that are presented in \cite{Azizi} satisfy the energy conditions. The modeling of wormholes in $f(R, T)$ gravity is presented in \cite{Moa} by Moraes and Sahoo. They have identified solutions with linear EoS and a variable EoS parameter that satisfy the energy conditions for the linear EoS model. Zubair et. al. discuss spherically symmetric wormhole solutions in $f(R, T)$ modified theory of gravity by introducing well-known non-commutative geometry in terms of Gaussian and Lorentzian distributions of string theory \cite{Zub}. Rosa and Kull have demonstrated that traversable wormhole solutions for the non-vanishing redshift function in the linear form of $f(R,T)=R + \lambda T$ gravity satisfy the  energy conditions for the entire spacetime \cite{Rosa}. Sharif and Nawazish explored wormhole solutions in spherically symmetric spacetime using the Noether symmetry approach within the framework of $f(R, T)$ gravity \cite{Shar}. In \cite{Shw}, two different traversable wormhole geometries, namely  exponential and  power law shape functions have been utilized to model the wormholes. In \cite{Cha}, three different models of $f(R, T)$ are investigated to find exact wormhole solutions.  Solutions for charged wormholes in  $f(R, T)$ extended theory of gravity  have been presented in \cite{Charge}. Sharif and Fatima have introduced traversable wormhole solutions through Karmarkar condition in $f(R, T)$ theory \cite{Sharif}. In \cite{Sah} a new hybrid shape function for wormhole in  modified $f(R, T)$ gravity is proposed. Many other wormhole solutions have been discussed in the context of $f(R, T)$ gravity \cite{Moa2, Zub2, fr1, fr2, fr3, fr4, Yousaf, squared, Sha, Sahoo, god, Bha, Ban, Mish, Tang, Noori, Sama, Gosh}. Some of the solutions in the context of $f(R, T)$ satisfy the ECs  while others do not respect the energy conditions. Despite the numerous studies on wormholes in the context of  $f(R, T)$, non-exotic wormhole solutions for $f(R,T)=R+2\lambda T$ are scarce in the existing literature. Wormhole with linear EoS and constant redshift function are presented in \cite{Azizi} and \cite{Moa}. In existing literature, solutions with constant redshift function that satisfy the ECs are  limited to a power-law shape function and linear EoS for $f(R, T)=R+\lambda T$ \cite{Azizi, Moa}. In this paper,  we introduce a diverse range of solutions within the framework of  $f(R, T)=R+2\lambda T$ that also adhere to the energy conditions. We apply the approach outlined in reference \cite{variable} to construct asymptotically flat wormholes with a variable EoS parameter.

Our article is organized as follows: Sec. \ref{sec2}, delves into the criteria and equations that govern wormholes. Subsequently, we provide a concise overview of $f(R, T)$ theory and the classical ECs.  In Sec. \ref{sec3}, we explore various shape functions within the realm of $f(R, T)$ gravity. We present solutions which satisfy the ECs in Sec.  \ref{sec4}. The physical properties of the solutions are presented in this section. Finally, we offer our concluding remarks in the last section.  In this paper, We have assumed gravitational units, i.e., $c = 8 \pi G = 1$.

\section{Basic formulation of wormhole } \label{sec2}
In this section, we introduce the basic structure of the wormhole theory and a brief review of $f(R, T)$ gravity formalism. Typically, the line element of a static and spherically symmetric wormhole is given by
\begin{equation}\label{1}
ds^2=-U(r)dt^2+\frac{dr^2}{1-\frac{b(r)}{r}}+r^2(d\theta^2+\sin^2\theta,
d\phi^2)
\end{equation}
where $U(r)=\exp (2\phi(r))$. The function $b(r)$ is called the shape function and $ \phi(r)$ is called the redshift function which can be used to detect the redshift of the signal  by a distance observer. The shape function should obey
\begin{equation}\label{2}
b(r_0)=r_0.
\end{equation}
where  $r_0$ is the wormhole throat.  Additionally, two other conditions must be met to ensure the existence of a traversable wormhole,
\begin{equation}\label{3}
b'(r_0)<1
\end{equation}
and
\begin{equation}\label{4}
b(r)<r,\ \ {\rm for} \ \ r>r_0.
\end{equation}
Equation (\ref{3}) is well-known as the flaring-out condition which leads to violation of NEC in the background of GR.
Asymptotically flat condition can be formulated as follows
\begin{equation}\label{5}
\lim_{r\rightarrow \infty}\frac{b(r)}{r}=0,\qquad   \lim_{r\rightarrow \infty}U(r)=1.
\end{equation}
It is necessary that redshift function be finite everywhere to avoid the existence of the horizon. To simplify, we have focused on solutions with a constant redshift function.
In this article, we consider an anisotropic fluid in the form  of  $T^{\mu}_{\nu}=diag[-\rho, p,p_t,p_t]$, where $\rho$ is the energy density, $p$ is the
 radial pressure and $p_t$ denotes the tangential pressure, respectively.

Let us briefly review the $f(R, T)$ formalism. The total action for $f(R, T)$ gravity takes the form:
\begin{equation}\label{6}
S=\int\frac{1}{2}f(R,T)\sqrt{-g}\;d^{4}x+\int L_{m}\sqrt{-g}\;d^{4}x.
\end{equation}
where $f(R, T)$ is a general function  of $R$ (Ricci scalar) and $T$ (trace of the energy-momentum tensor), $g$ is the determinant
of the metric, and $L_m$ is the matter Lagrangian density. The connection between $L_m$ and the energy-momentum tensor is given by
\begin{equation}\label{7}
T_{ij}= - \frac{2}{\sqrt{-g}}\left[\frac{\partial(\sqrt{-g}L_{m})}{\partial g^{ij}}-\frac{\partial}{\partial x^{k}}\frac{\partial(\sqrt{-g}L_m)}{\partial(\partial g^{ij}/\partial x^{k})}\right].
\end{equation}
In most of the papers in $f(R, T)$ gravity, $L_m$ is assumed to be dependent only on the metric component not on its derivatives. We also use this assumption, so
\begin{equation}\label{8}
T_{ij}= g_{ij}L_{m} - 2\frac{\partial L_{m}}{\partial g^{ij}}.
\end{equation}
One can vary the action \ref{6} with respect to metric to find
\begin{multline}\label{9}
f_R(R,T)\left(R_{ij}-\frac{1}{3} Rg_{ij}\right) + \frac{1}{6}f(R,T)g_{ij} \\= \left(T_{ij}-\frac{1}{3}Tg_{ij}\right)-f_T(R,T)\left(T_{ij} -\frac{1}{3}Tg_{ij}\right)\\-f_T(R,T)\left(\theta_{ij}-\frac{1}{3}\theta g_{ij}\right)+\nabla_i\nabla_jf_R(R,T),
\end{multline}
 where $f_R (R,T)\equiv \frac{\partial f(R,T)}{\partial R}$, $f_T (R,T)\equiv \frac{\partial f(R,T)}{\partial T}$ and
 \begin{equation}\label{10}
\theta_{ij}=g^{ij}\frac{\partial T_{ij}}{\partial g^{ij}}.
\end{equation}
There are numerous options for Lagrangian density, such as  $L_m = T$, $L_m =P$ where $P=\frac{p_r+2p_t}{3}$, and $L_m =-\rho$  with different physical interpretation. We assume $L_m =-\rho$ which is a natural choice. It should be noted that $L_m = -\rho$ leads to non-vanishing extra force. Hence, Eq.(\ref{10}) give
\begin{equation}\label{11}
\theta_{ij}=-2T_{ij}-\rho g_{ij}.
\end{equation}
There are many models in the general forms
\begin{equation}\label{13}
f(R,T)=f_1(R)+f_2(T),
\end{equation}
\begin{equation}\label{13a}
f(R,T)=f_1(R)+f_2(R)f_3(T).
\end{equation}
The simplest scenario is as follows
\begin{equation}\label{14}
f(R,T)=R+2\lambda T.
\end{equation}
Using (\ref{1}), (\ref{9}) and (\ref{13}), one can find the following field equations
\begin{equation}\label{15}
\frac{b'}{r^2}=(1+\lambda)\rho-\lambda(p_r+2p_l),
\end{equation}
\begin{equation}\label{16}
-\frac{b}{r^3}= \lambda \rho+(1+3\lambda)p_r+2\lambda p_l,
\end{equation}
\begin{equation}\label{17}
\frac{b-b'r}{2r^3}= \lambda \rho+\lambda p_r+(1+4\lambda) p_l.
\end{equation}
which the prime denotes the derivative $\frac{d}{dr}$.
The field equations stated above admit the solutions
\begin{equation}\label{18}
\rho=\frac{b'}{r^2(1+2\lambda)},
\end{equation}
\begin{equation}\label{19}
p_r= - \frac{b}{r^3(1+2\lambda)},
\end{equation}
\begin{equation}\label{20}
p_l= \frac{b-b'r}{2r^3(1+2\lambda)}.
\end{equation}
It is worth mentioning that the field equations changed only in Einstein's gravitational constant.

It has been demonstrated that wormhole solutions within ordinary GR entail the violation of ECs.  To ensure a positive stress-energy tensor in the existence of matter, the ECs present feasible methods. These ECs, known as the NEC, dominant energy condition (DEC), weak energy condition (WEC), and strong energy condition (SEC), are specifically defined to facilitate the achievement of this desired state.
\begin{eqnarray}\label{21}
\textbf{NEC}&:& \rho+p\geq 0,\quad \rho+p_t\geq 0 \\
\label{21a}
\textbf{WEC}&:& \rho\geq 0, \rho+p\geq 0,\quad \rho+p_t\geq 0, \\
\textbf{DEC}&:& \rho\geq 0, \rho-|p|\geq 0,\quad \rho-|p_t|\geq 0, \\
\textbf{SEC}&:& \rho+p\geq 0,\, \rho+p_t\geq 0,\rho+p+2p_t \geq 0. \label{21b}
\end{eqnarray}
 Derived from the Raychaudhury equations, these conditions serve as indispensable tools for understanding the geodesics of the Universe. By utilizing these equations, we are able to unravel the complex paths followed by cosmic objects. Now, according to \cite{fq}, by defining the functions,
\begin{eqnarray}\label{22}
 H(r)&=& \rho+p ,\, H_1(r)= \rho+p_t,\, H_2(r)= \rho-|p|, \nonumber \\
 H_3(r)&=&\rho-|p_t|,\, H_4(r)= \rho+p+2p_t .
\end{eqnarray}
We can investigate the ECs in the recent part of this paper. It was mentioned that wormholes in the context of $f(R, T)$ have been studied extensively in the literature. Solutions with $f(R, T)$ in the form of (\ref{14}) are explored in \cite{Azizi} and \cite{Moa}. It was shown that $H(r)>0$ and flaring out condition leads to
\begin{equation}\label{23}
\lambda < -\frac{1}{2}
\end{equation}
Although solutions with linear EoS
 \begin{equation}\label{24}
p(r)=w\rho
\end{equation}
 are explored in \cite{Azizi} and\cite{Moa} but as we know, there are a few  solutions for $f(R, T)=R+2\lambda T$, which satisfy the ECs in the existing literature. In this paper, we present other classes of solutions that satisfy ECs. For the simplicity, we set $r_0=1$ in the latter part of this paper.

\section{Wormhole solutions }\label{sec3}

 Exotic matter is the primary motivation for exploring wormholes in the context of modified gravities. Various algorithms are available to study wormholes within the framework of $f(R, T)$ gravity. For instance,  one can use an EoS and field equations to find exact wormhole solutions \cite{Azizi, Moa, fr1}. In another algorithm, the EoS can be not imposed, then we can use different  shape functions with free parameters to find the desired solutions \cite{Rosa, Shw}. Solutions where the energy density is a function of  $R$ and $R'$ are discussed in \cite{fr3}. In \cite{Zub}, energy density was assumed to be a function of the radial coordinate, and then $b(r)$ was presented. However, the solutions presented in \cite{Zub} do not adhere to the energy conditions.
  In \cite{Azizi}, wormhole with vanishing redshift function and $f(R, T)$ in the form of (\ref{14}) has been investigated. They have shown that  barotropic EoS (\ref{24}) leads to
 \begin{equation}\label{24a}
b(r)=r^{1/\omega}.
\end{equation}
This class of solutions with conditions (\ref{23}) and $\omega\leq-1$ satisfy the ECs.

It should be noted that Eqs.(\ref{18}-\ref{20}) give
 \begin{equation}\label{24b}
\rho+p+2p_t=0.
\end{equation}
This equation demonstrates that solutions satisfying NEC also comply with SEC. It is a crucial point to keep in mind while exploring wormhole solutions.
Considering Eq.(\ref{24b}) leads to an EoS for $p_t$ in term of $p$. As an illustration, $p=\omega(r) \rho$ results in
 \begin{equation}\label{24c}
p_t=-\frac{1+\omega(r)}{2}\rho,
\end{equation}
indicating the coexistence of linear EoS for lateral and radial pressure. One can note that for linear EoS,
 \begin{equation}\label{24cc}
p_t=\gamma p
\end{equation}
where
 \begin{equation}\label{24d}
 \gamma=-\frac{1+\omega}{2\omega}
\end{equation}
This equation implies that linear EoS for radial pressure leads to linear EoS for lateral pressure.
\begin{figure}
\centering
  \includegraphics[width=3 in]{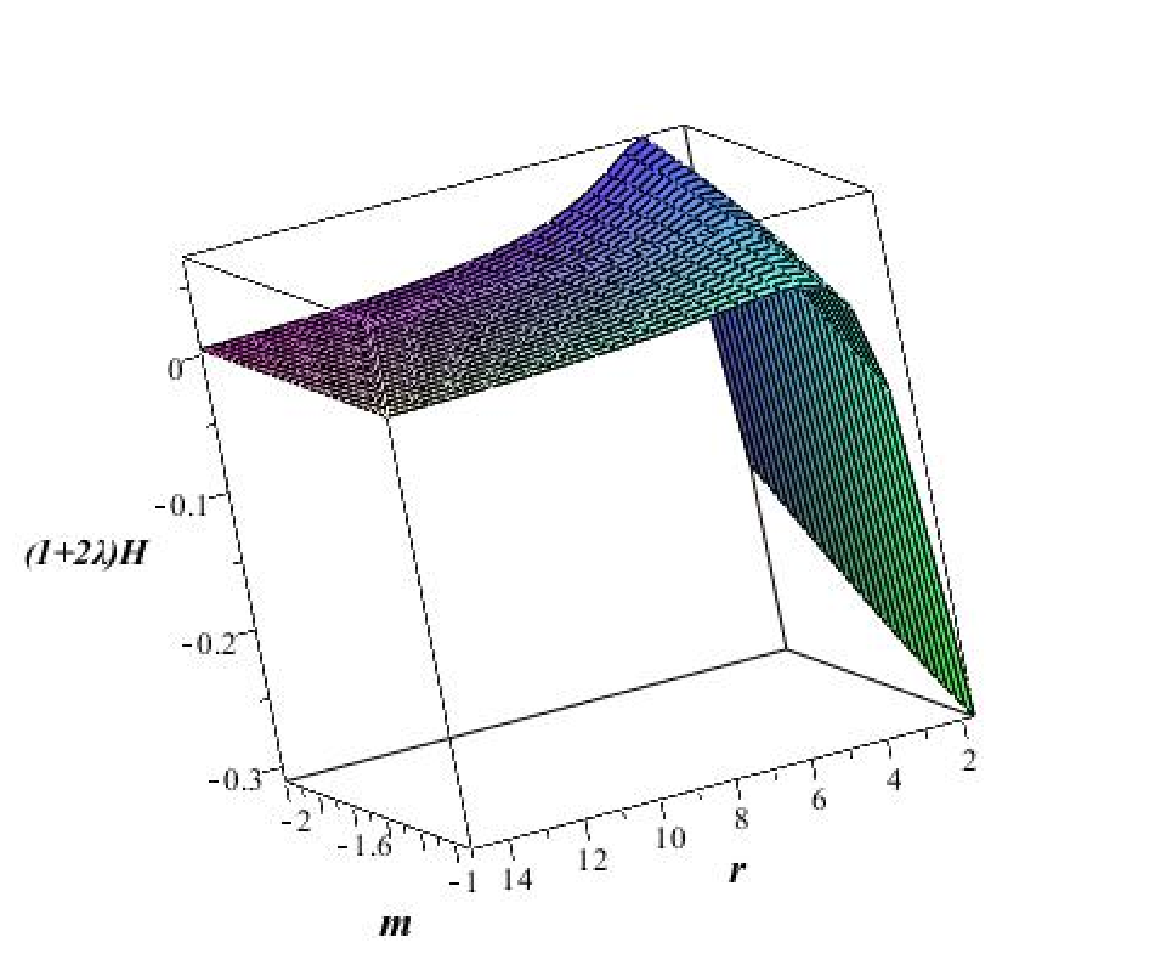}
\caption{The figure represents the $(1+2\lambda)H(r,m)$ against $r$ and $m$ for $A=2$,  which is  negative in some range. See the text for details.}
 \label{fig1a}
\end{figure}

Now, we examine  some  choices for shape function to find exact wormhole solutions. To this end, well-known shape functions from the literature are employed.
 Our options are as follows
 \begin{eqnarray}\label{25}
b(r)=Ar^m+(1-A),
\end{eqnarray}
 \begin{equation}\label{26}
b(r)=\frac{\tanh(r)}{\tanh(1)},
\end{equation}
\begin{equation}\label{27a}
b(r)=c \ln(r)+1,
\end{equation}
\begin{equation}\label{27}
b(r)=\frac{a^r}{a}.
\end{equation}
These shape functions have been extensively studied in the current literature regarding different theories of gravity. They are consistent with Eqs.(\ref{2}-\ref{5}). In this work, our emphasis is on solutions with positive energy density, which seems to be more viable. The shape function (\ref{25}) is presented in \cite{phantom1} to study asymptotically flat wormhole solutions with phantom EoS. This shape function  gives
\begin{equation}\label{28}
\rho(r)=\frac{Am}{1+2\lambda}r^{m-3}.
\end{equation}
To have a positive energy density with $\lambda < -\frac{1}{2}$,
\begin{equation}\label{28a}
Am<0
\end{equation}
is necessary. One can easily show
\begin{equation}\label{1a}
\frac{H(r,m,A)}{1+2\lambda}=\frac{(Amr^{m-1}-Ar^m-1+A)}{r^3},
\end{equation}
\begin{equation}\label{2a}
\frac{H_1(r,m,A)}{1+2\lambda}=\frac{(Amr^{m-1}+Ar^m+1-A)}{2r^3}.
\end{equation}
so
\begin{equation}\label{3a}
\lim_{r \rightarrow r0} H(r,m,A)(1+2\lambda)=-1+Am,
\end{equation}
\begin{equation}\label{4a}
\lim_{r \rightarrow r0} H_1 (r,m,A)(1+2\lambda)=\frac{1+Am}{2},.
\end{equation}
\begin{figure}
\centering
  \includegraphics[width=3 in]{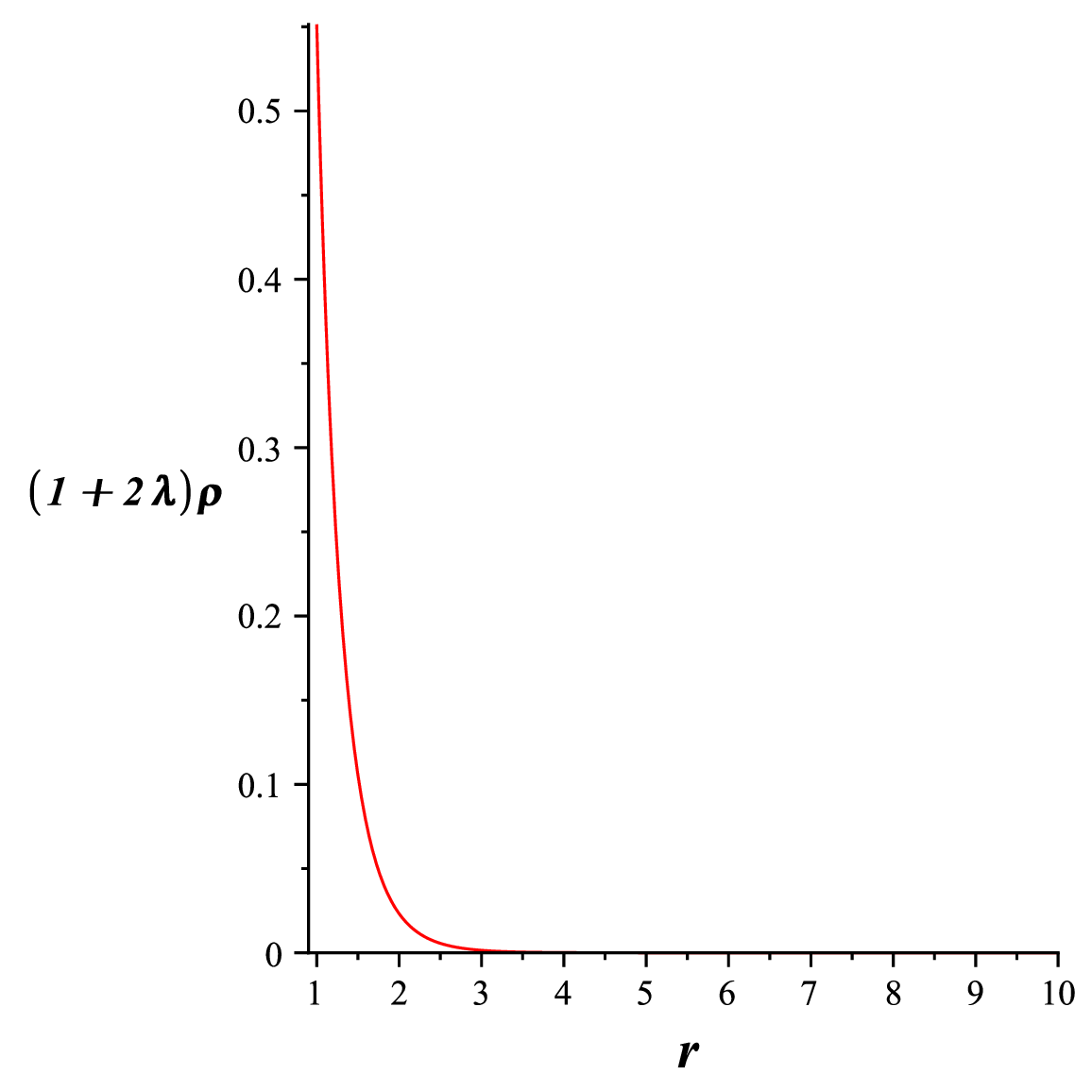}
\caption{The figure represents the $(1+2\lambda)\rho(r,n)$ against radial coordinate,  which is  positive in the whole range. See the text for details.}
 \label{fig1}
\end{figure}
The conditions (\ref{3a}) and (\ref{4a}) are satisfied when  $Am<-1$. The overall behavior of  $H(r,m,A)(1+2\lambda)$  is intricate, but we have plotted $H(r,m)(1+2\lambda)$ as a function of $r$ and $m$ for $A=2$ in Fig,(\ref{fig1a}). This figure verifies the violation of ECs within a certain range of the radial coordinate. This procedure can be used for the other values of $A$ and $m$. Now, we can conclude that the shape function (\ref{25}) does not satisfy the ECs in the context of $f(R,T)$ theory of gravity.

The shape function (\ref{26}) has been investigated in \cite{Lyra} in the context of Lyra manifold which yields
\begin{equation}\label{28b}
\rho(r)=\frac{1-\tanh(r)^2}{(1+2\lambda)\tanh(1)r^2},.
\end{equation}
We have plotted $(1+2\lambda)\rho(r)$ against $r$ in Fig.(\ref{fig1}). This figure guarantees that
 the energy density is negative  everywhere for $\lambda<-1/2$. Generally, Eq.(\ref{18}) demonstrates that solutions with positive energy density in ordinary GR give a negative energy density for $1+2\lambda<0$ in the background of $f(R, T)$.

\begin{figure}
\centering
  \includegraphics[width=3 in]{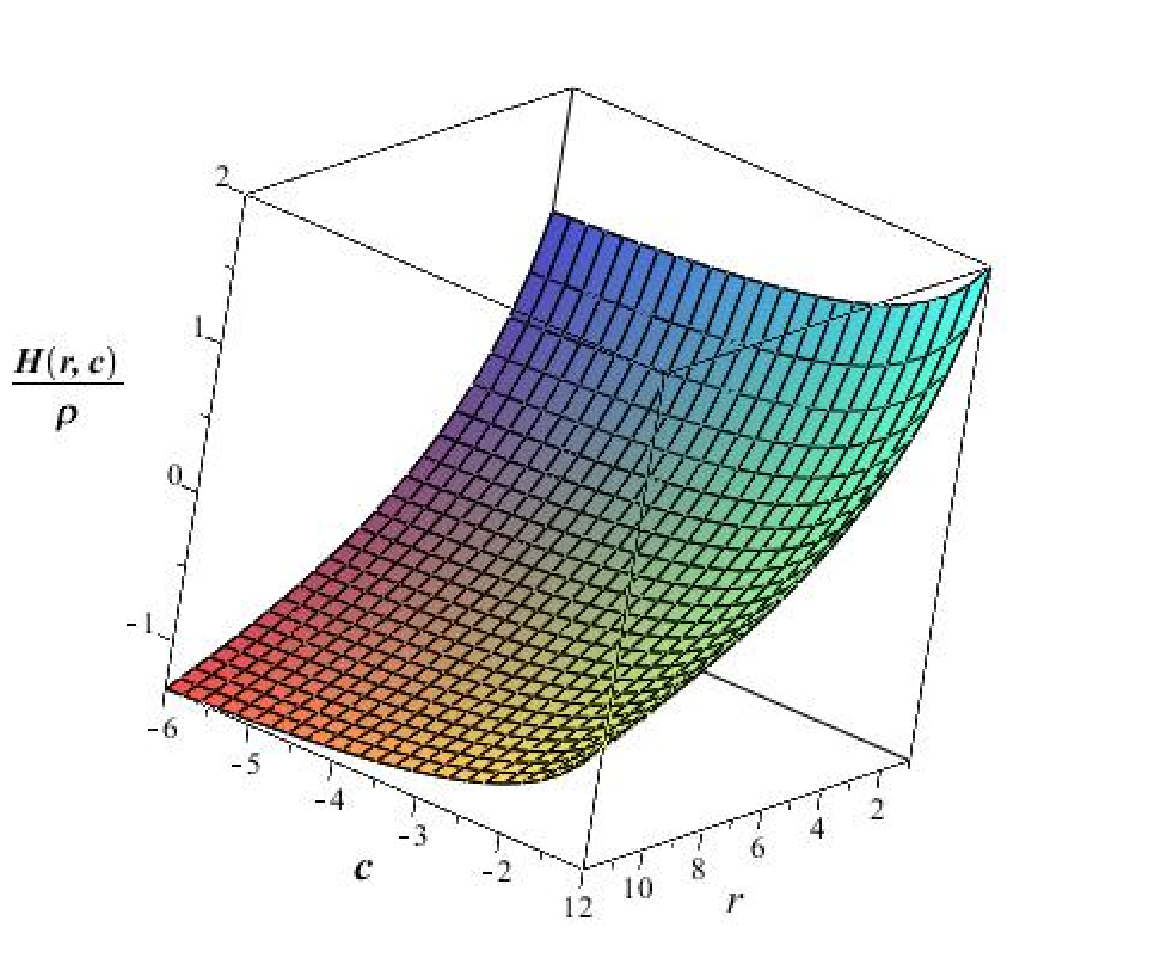}
\caption{The figure represents the $\frac{H(r,c)}{\rho}$ against radial coordinate and $c$,  which is  negative in  some range. See the text for details.}
 \label{fig2}
\end{figure}

It is easy to show that the energy density related to the shape function (\ref{27a}) is
\begin{equation}\label{8b1}
\rho(r)=\frac{c}{(1+2\lambda)r^3}.
\end{equation}
Since $1+2\lambda<0$ is assumed, the energy density is positive for $c<0$. This shape function provides
\begin{equation}\label{8b2}
\frac{H(r,c)}{\rho}=1-\frac{c\ln(r)+1}{c},
\end{equation}
 \begin{equation}\label{8b3}
\frac{H_1(r,c)}{\rho}=1+\frac{c\ln(r)+1}{c}.
\end{equation}
We have plotted $\frac{H(r,c)}{\rho}$ as a function of $r$ and $c$ in Fig.(\ref{fig2}). This figure demonstrates that the NEC is not valid for this shape function.

\begin{figure}
\centering
  \includegraphics[width= 3 in]{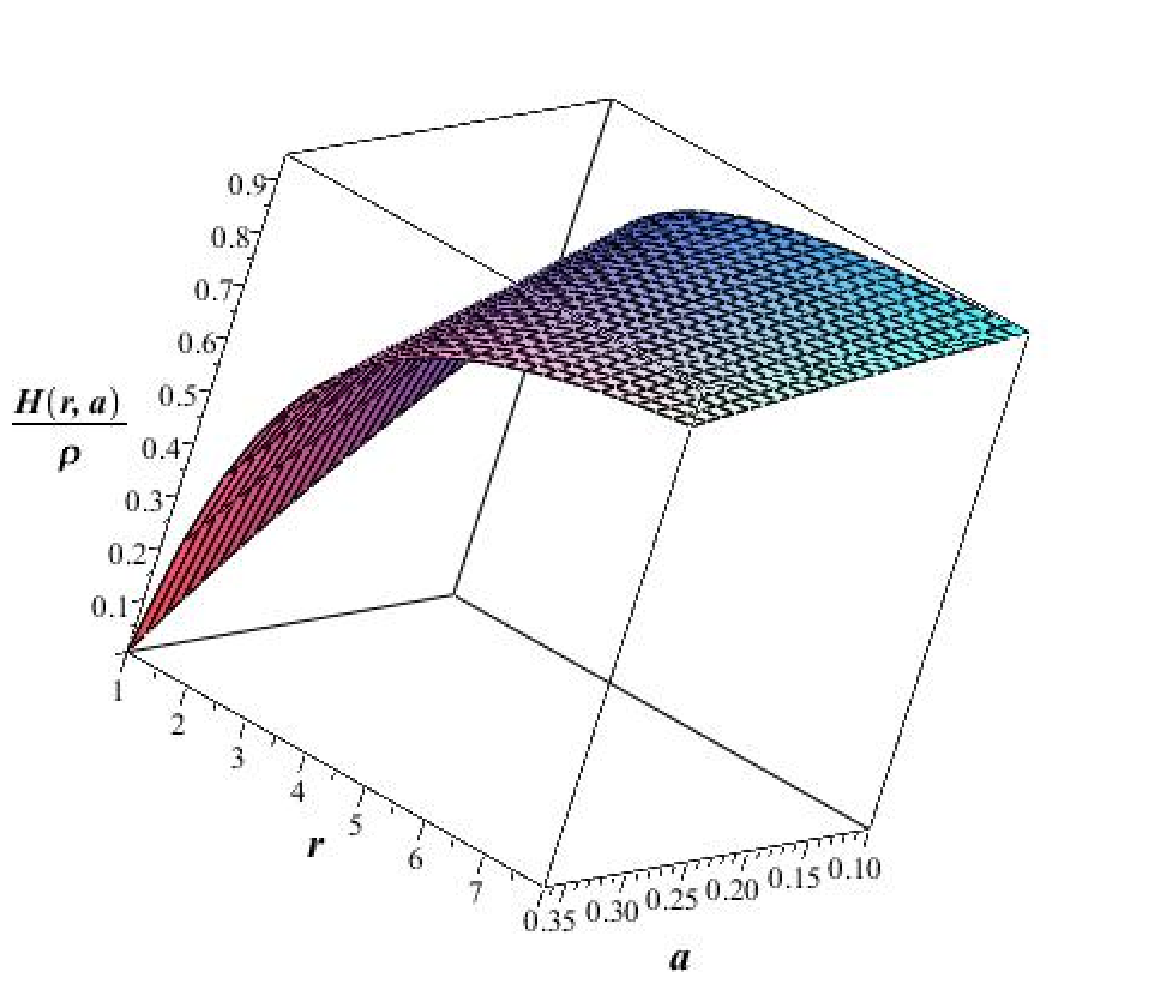}
\caption{The figure represents the $\frac{H(r,c)}{\rho}$ against radial coordinate and $a$ for $0<a<1/e$,  which is  positive in the whole range. See the text for details.}
 \label{fig3}
\end{figure}

\begin{figure}
\centering
  \includegraphics[width=3 in]{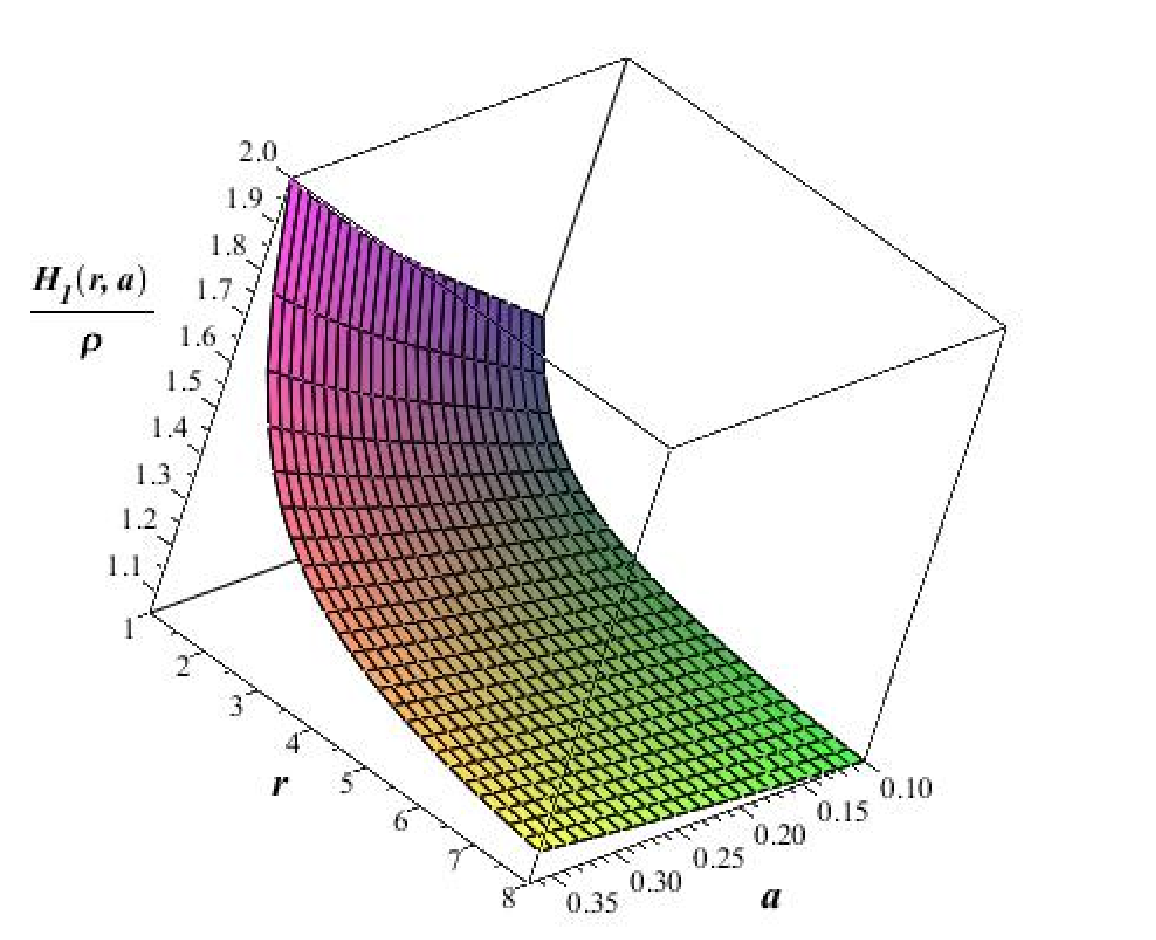}
\caption{The figure represents the $\frac{H_1(r,a)}{\rho}$ against radial coordinate and $a$ for $0<a<1/e$, which is positive in the whole range. See the text for details.}
 \label{fig4}
\end{figure}

The shape function (\ref{27}) admits the conditions (\ref{1}-\ref{5}) for $0<a<1$. This shape function has been investigated in \cite{Shw} in the framework of $f(R, T)=R+2\lambda T$. Using (\ref{27}) in (\ref{18}) gives
\begin{equation}\label{28b4}
\rho(r)=\frac{\ln(a)}{a(1+2\lambda)}\frac{a^r}{r^2},
\end{equation}
which is positive for $0<a<1$ and $\lambda<-1/2$. The related $H(r,a)/\rho$ and $H2(r,a)/\rho$ are
\begin{equation}\label{28b5}
\frac{H(r,a)}{\rho}=1-\frac{1}{\ln(a)r}
\end{equation}
and
 \begin{equation}\label{28b6}
\frac{H_1(r,a)}{\rho}=1+\frac{1}{\ln(a)r}.
\end{equation}
It is easy to show that
\begin{equation}\label{28b7}
\lim_{r\longrightarrow 0}\frac{H_1(r,a)}{\rho}=1+\frac{1}{\ln(a)}.
\end{equation}
 which verifies the violation of NEC at the wormhole throat for $1/e<a<1$.
 We have plotted $\frac{H(r,a)}{\rho}$ and $\frac{H_1(r,a)}{\rho}$as a function of $r$ and $a$ in Figs.(\ref{fig3}) and (\ref{fig4}) for  $0<a<1/e$, which are positive in the entire range. It is easy to show that $\frac{H_2(r,c)}{\rho}=\frac{H(r,c)}{\rho}$ and $\frac{H_3(r,a)}{\rho}=\frac{H_1(r,c)}{\rho}$. Thus, one can conclude that, for  $0<a<1/e$, the shape function (\ref{27}) satisfies all the ECs.

Elizalde and Khurshudyan have studied solutions with varying Chaplygin gas and barotropic fluid in the context of $f(R,T)$\cite{fr2}. They have identified shape functions for certain specific EoS that do not adhere to the ECs.   Additionally, some of these shape functions do not exhibit asymptotically flat behavior, particularly for EoS in the following form:
\begin{equation}\label{28c}
p=\omega (r) \rho,
\end{equation}
where
\begin{equation}\label{28d}
\omega(r)=\omega b(r)^\nu  \rho(r).
\end{equation}
The shape function
\begin{equation}\label{28e}
b(r)=(1-\frac{\nu \log(r)}{\omega})^{1/\nu}.
\end{equation}
is achieved. This shape function does not respect the ECs.
In \cite{Moa}, it was shown that  EoS in the form (\ref{28c}), with $\omega(r)=Br^m$ leads to
\begin{equation}\label{28f}
b(r)=\exp(c-\frac{1}{Bmr^m}).
\end{equation}
Moares and Sahoo explained that this shape function violates the ECs for $m>0$ in the background of $f(R,T)$.
Most of the shape functions in the current literature are unable to address the issue of exotic matter in the context of $f(R,T)$ gravity as described in  (\ref{14}). We propose an algorithm to discover asymptotically flat wormhole solutions within the framework of $f(R,T)=R+2\lambda T$. Previous studies on solutions with varying EoS parameter in the context of $f(R,T)$ theory of gravity are found in \cite{Moa} and \cite{fr1, fr2, fr3} but these solutions violate the ECs. In the next sections, by using a variable EoS parameter, we will seek solutions that satisfy the ECs. We start with a particular form of asymptotically linear EoS to construct the desired wormhole solutions, then we will study the physical properties of the solutions.

\section{Solutions satisfying energy conditions }\label{sec4}
In the previous section, it was demonstrated that most known wormhole solutions within the framework of $f(R,T)=R+2\lambda T$ violate the ECs. In this section, we aim to discover asymptotically flat wormhole solutions that adhere to the energy conditions in the context of $f(R,T)$.  In \cite{variable}, we have shown that variable EoS in the context of GR results in a small amount of exotic matter to construct wormhole solutions.
 We apply the same approach to identify exact wormhole solutions within the  background of $f(R,T)$. We adopt a linear-like EoS given by
\begin{equation}\label{f3}
p=\omega_{eff}(r)\rho(r)=(\omega_\infty+g(r))\rho(r),
\end{equation}
where $\omega_{eff}(r)$ is the effective state parameter and  $\omega_\infty$, denotes the constant state parameter at the large radial coordinate.
To ensure an asymptotically linear EoS, we impose the condition
\begin{equation}\label{f4}
\lim_{r\rightarrow \infty}g(r)=0,
\end{equation}
to achieve this characteristic.  Through the selection of various forms of the $g(r)$ function and  fine-tuning the free parameters, we present solutions that satisfy the ECs. Using (\ref{18}), (\ref{19}) and (\ref{f3}), we can obtain
\begin{equation}\label{29}
b(r)=\exp(\int \frac{dr}{r\omega_\infty +rg(r)}).
\end{equation}
Many broad families of solutions for different $g(r)$ functions can be found. In the reverse approach, we have the flexibility to arbitrarily choose the shape function and subsequently find the EoS parameter.
Now, we proceed to verify the energy conditions for this particular class of solutions. For $\rho>0$, condition $H\geq 0$ gives
\begin{equation}\label{m}
-1\leq \omega(r),
\end{equation}
also $H_1\geq 0$ and Eq.(\ref{24c}) indicates that
\begin{equation}\label{m1}
 \omega(r)\leq1,
\end{equation}
 so
\begin{equation}\label{29a1}
-1\leq \omega(r)\leq1,
\end{equation}
is an essential requirement for the maintenance of non-exotic wormholes. It is straightforward to show that $H_2\geq0$ and $H_3\geq0$ are satisfied when Eq.(\ref{29a1}) is valid. Thus, equation ( \ref{24b}) demonstrates that all the ECs are satisfied when condition (\ref{29a1}) is valid.

\subsection{ Solutions with $\omega_\infty \neq 0$ }\label{subsec1}
In this subsection, we assume $\omega_\infty \neq 0$. Let us study some different $g(r)$ models to establish traversable wormholes.
As our initial model,  we consider
 \begin{equation}\label{29a11}
  g(r)=\frac{1}{r},
  \end{equation}
 which results in
\begin{equation}\label{29a}
b(r)=c(\omega_\infty r+1)^{-1/\omega_\infty},
\end{equation}
where $c$ is the constant of integration. One can find $c=(\omega_\infty +1)^{1/\omega_\infty}$ by using the condition (\ref{2}).

At a large distance from the throat, the condition (\ref{29a1}) changes to
\begin{equation}\label{29a2}
-1\leq \omega_\infty\leq1.
\end{equation}
 On the other hand, at the throat, condition (\ref{29a1}) gives
\begin{equation}\label{29a3}
-2\leq \omega_\infty\leq 0.
\end{equation}
Considering the conditions (\ref{29a2}) and (\ref{29a3}) implies that
\begin{equation}\label{9a}
-1<\omega_\infty<1
\end{equation}
 must be satisfied to have non-exotic wormhole solutions. In this work, we are interested in static and spherically symmetric asymptotically flat traversable wormhole solutions. Thus, Eq.(\ref{29a}) leads to
\begin{equation}\label{29a4}
\frac{-1}{\omega_\infty}\leq 1.
\end{equation}
It is evident that conditions (\ref{29a3}) and (\ref{9a}) are not compatible. Therefore, it can be inferred that $g(r)=\frac{1}{r}$ does not lead to solutions that meet the ECs.

\begin{figure}
\centering
    \includegraphics[width=3 in]{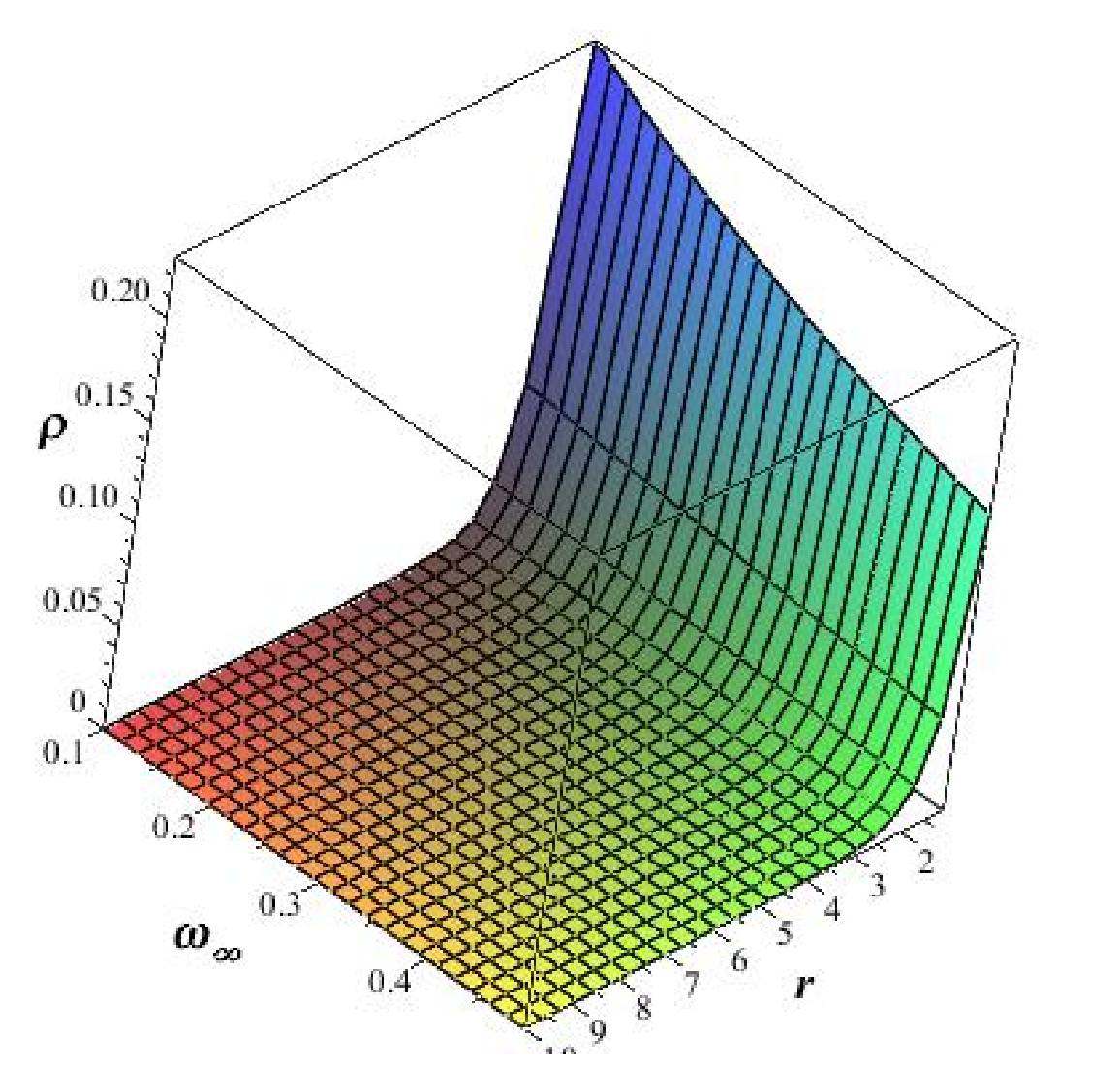}
\caption{The graphical behavior of $\rho$,  against $r$  and $\omega_\infty$ for the case $\lambda=-3$ and $D=\frac{1}{2}$. It is clear that this function is positive in the entire range of $r$ and $0<\omega_\infty<\frac{1}{2}$ . See the text for details.}
 \label{fig5}
\end{figure}

Let us explore another example of wormhole solutions. The function
\begin{equation}\label{29bb}
 g(r)=\frac{D}{r},
 \end{equation}
where $D>0$, is the next potential candidate that yields
\begin{equation}\label{29b}
b(r)=(\omega_\infty +D)^{1/\omega_\infty}(\omega_\infty r+D)^{-1/\omega_\infty}
\end{equation}
The validity of  condition (\ref{29a1}) at the throat and large distance implies that
\begin{equation}\label{2b}
-1\leq \omega_\infty\leq 1-D.
\end{equation}
 Applying the asymptotically flat condition and (\ref{2b}) indicates that
\begin{equation}\label{29bb1}
0\leq \omega_\infty\leq 1-D.
\end{equation}
It is worth mentioning that Eq. (\ref{29bb1}) is  constrained by the value of $D$. The subsequent step involves verifying the positivity of the energy density. Equation (\ref{18}) for the shape function (\ref{29b}) illustrates that
\begin{equation}\label{b3}
\rho(r)=-\frac{(\omega_\infty+D)^{1/\omega_\infty}}{1+2\lambda}\frac{1}{r^2(\omega_\infty r+D)^{1+\omega_\infty}}.
\end{equation}
This equation clarifies that
\begin{equation}\label{29b2}
 \omega_\infty \geq-D
\end{equation}
must be satisfied to ensure a positive energy density. As an example, we set $D= 1/ 2$ which shows that
\begin{equation}\label{29b1}
0\leq \omega_\infty\leq \frac{1}{2}.
\end{equation}
The energy density is plotted in Fig.(\ref{fig5}) for $\lambda=-3$ and $D=1/2$ which is positive in the entire range of $r$ and $0<\omega_\infty<1/2$. Thus this class of solutions satisfies the ECs.

Let us explore the connections between  the parameters in the shape function and physical parameters. The boundary limits explain that
\begin{equation}\label{b4}
 \omega_0=\frac{p_0}{\rho_0}=\lim_{r\longrightarrow r_0}\omega(r) = \omega_\infty+D
\end{equation}
so
\begin{equation}\label{b5}
 D=\omega_0- \omega_\infty.
\end{equation}
The energy density at the wormhole throat is given by
\begin{equation}\label{b6}
 \rho_0=\lim_{r\longrightarrow r_0}\rho(r) = -\frac{(\omega_\infty+D)^{-1}}{1+2\lambda}
\end{equation}
This equation along with  Eq.(\ref{b5}) show that
\begin{equation}\label{b7}
 \omega_0=-\frac{(\rho_0)^{-1}}{(1+2\lambda)}.
\end{equation}
Using Eq.(\ref{b7}) indicates
\begin{equation}\label{b8}
p_0= \omega_0\rho_0=-\frac{1}{(1+2\lambda)}.
\end{equation}
Of course, this equation can be archived by using (\ref{2}) and (\ref{19}) for a general shape function.
One can utilize (\ref{b8}) to establish the relationship between $\lambda$ and radial pressure at the throat as follows
\begin{equation}\label{b9}
\lambda=\frac{1}2{p_0}-\frac{1}{2}.
\end{equation}
As it was mentioned, $\lambda<-\frac{1}{2}$ is the necessary condition to achieve solutions in the context of $f(R,T)=R+2\lambda T$. Thus, Eq.(\ref{b9}) explains that $p_0<0$ is the essential condition to sustain non-exotic wormhole solutions.
\begin{figure}
\centering
  \includegraphics[width=3 in]{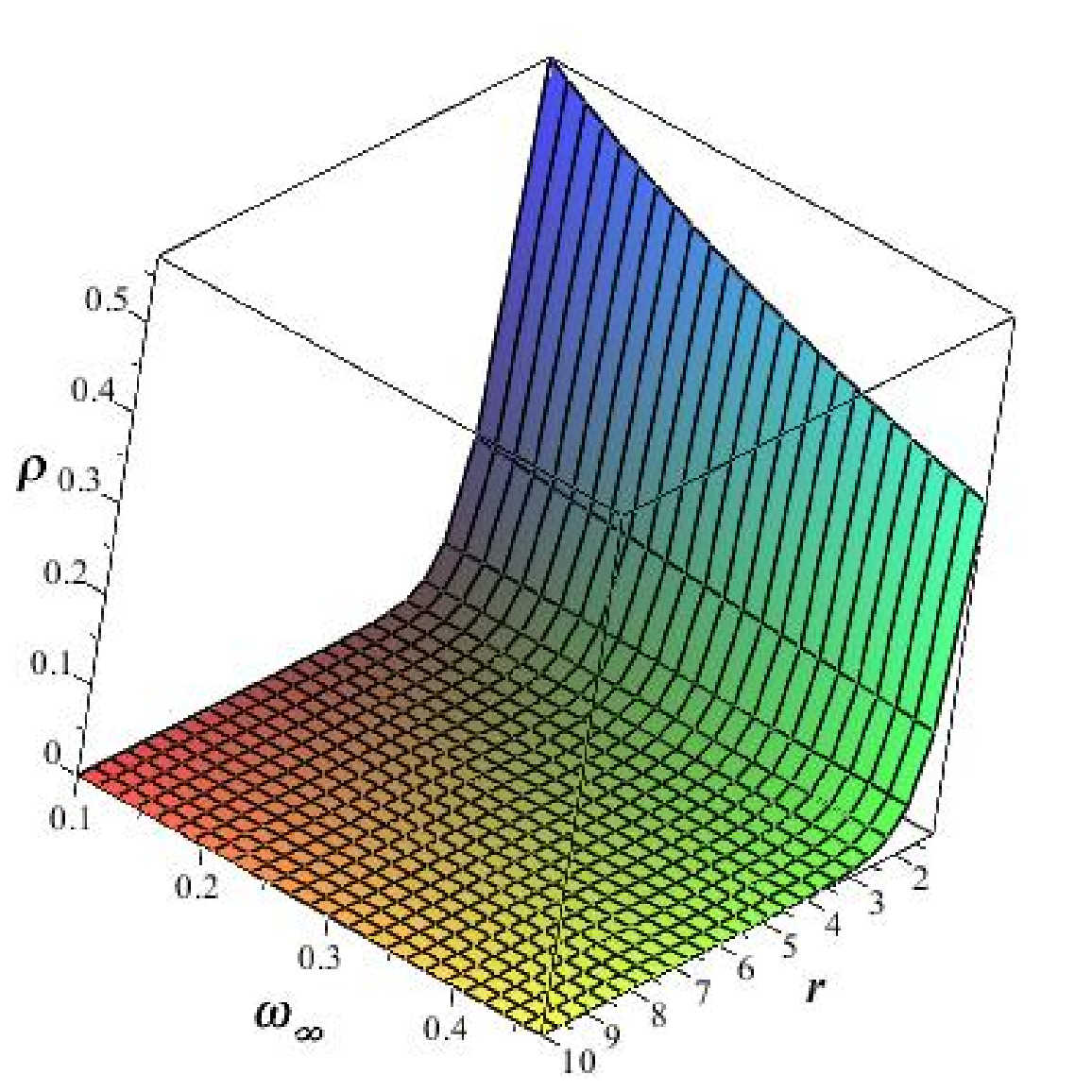}
\caption{The plot depicts the  behavior of $\rho(r,\omega_\infty)$ against $r$ and $\omega_\infty$ for the case $D=\frac{1}{2}$, $A=1$, and $\lambda=-3$. One can see that $\rho$ is a positive function in the entire range of $r$ and $0<\omega_\infty<\frac{1}{2}$ . See the text for details.}
 \label{fig6}
\end{figure}

The function
\begin{equation}\label{b10}
 g(r)=\frac{D}{r^n}
 \end{equation}
 is the next choice resulting in
\begin{equation}\label{29c}
b(r)=(\omega_\infty +D)^{1/n\omega_\infty}(\omega_\infty r^n+D)^{-1/n\omega_\infty},
\end{equation}
and
\begin{equation}\label{c2}
\rho(r)=-\frac{(\omega_\infty+D)^{\frac{1}{n\omega_\infty}}}{(1+2\lambda)}\frac{(\omega_\infty r^n+D)^{-\frac{1}{n\omega_\infty}-1}}{r^{3-n}}.
\end{equation}
 Following some simple computations, we determine that
\begin{equation}\label{29c1}
0\leq \omega_\infty\leq 1-D,
\end{equation}
and $\omega_\infty>-D$ must be satisfied to maintain  asymptotically flat wormhole solutions without exotic matter.
 In this case, the energy density at the throat is
 \begin{equation}\label{c3}
 \rho_0=\lim_{r\longrightarrow r_0}\rho(r) = -\frac{(\omega_\infty+D)^{-1}}{1+2\lambda}
\end{equation}
Also, Eqs.(\ref{b5}) and (\ref{b8}) are valid for this case.

The fourth assumption for $g(r)$ is
\begin{equation}\label{9c1}
 g(r)=\frac{1}{\ln (r)}.
 \end{equation}
which leads to
\begin{equation}\label{9c2}
 b(r)=\frac{(\omega_\infty \ln(r)+1)^{\frac{1}{\omega_\infty^2}}}{r^{\frac{1}{\omega_\infty}}}.
 \end{equation}
The function (\ref{9c1}) diverges at the wormhole throat, thus  the last choice is considered as
 \begin{equation}\label{8c1}
  g(r)=\frac{D}{A+\ln (r)}.
\end{equation}
This function results in
\begin{equation}\label{29e}
b(r)= r^{-1/\omega_\infty} \frac{(\omega_\infty  A+D +\omega_\infty  \ln (r))^{D/(\omega_\infty)^2}}{(\omega_\infty  A+D)^{D/(\omega_\infty)^2}}.
\end{equation}
Equation (\ref{29a1}),  at the throat gives
\begin{equation}\label{29f}
-1-\frac{D}{A}\leq \omega_\infty\leq 1-\frac{D}{A}.
\end{equation}
 Asymptotically flat condition and Eqs.(\ref{29a1}) and (\ref{29f}) indicate that
\begin{equation}\label{29g}
0\leq \omega_\infty\leq 1-\frac{D}{A}
\end{equation}
is the essential condition to have non-exotic solutions. The energy density is given by
\begin{eqnarray}\label{22h}
\rho(r)=r^{-3-\frac{1}{\omega_\infty}}\frac{(\omega_\infty A+D+\omega_\infty \ln(r))^{\frac{D}{\omega^2_\infty}}}{\omega_\infty(1+2\lambda)(\omega_\infty A+D)^{\frac{D}{\omega^2_\infty}}} \nonumber \\
\times(\frac{D}{\omega_\infty A+D+\omega_\infty \ln(r)}-1) .
\end{eqnarray}
Energy density is a function of $r, A, D$, and $\omega_\infty$ so analyzing the general behavior of this function is complicated. As an example, we have plotted $\rho$ versus $r$ and  $\omega_\infty$ for  $A=1$ and $D=1/2$ in Fig.(\ref{fig6}). This figure shows that energy density is positive for $r\geq r_0$. Since Eq.(\ref{29a1}) is valid, all the ECs are satisfied.

We have presented a broad class of exact wormhole solutions  using the variable EoS algorithm. We have assumed $\omega_\infty \neq 0$ but the special case $\omega_\infty = 0$ yields additional solutions. We will study solutions related to $\omega_\infty =0$ in the next section.

\subsection{Asymptotically Dust Solutions }\label{subsec2}
A special case for the EoS in the form of (\ref{f3}) can be considered for a vanishing $\omega_\infty$.These solutions can be classified as asymptotic dust solutions.  In this subsection,  we analyze these solutions for the previously defined $g(r)$ functions.  For  $g(r)=1/r$ and $\omega_\infty=0$, the shape function is given by
\begin{equation}\label{30}
b(r)=c \exp(-\frac{r^n}{nD})
\end{equation}
where $c$ is a constant of integration. Using (\ref{2}), we obtain
\begin{equation}\label{31}
b(r)= \exp(\frac{1}{nD}-\frac{r^n}{nD})
\end{equation}
This shape function is asymptotically flat and satisfies all the necessary conditions to sustain a traversable wormhole.
The corresponding energy density is
  \begin{equation}\label{32}
\rho(r)=-\frac{r^{n-3}\exp(\frac{1}{nD}-\frac{r^n}{nD})}{(1+2\lambda)D},
\end{equation}
Since $1+2\lambda <0$, it can be concluded that $\rho$ is positive for $D>0$. It should be noted that
\begin{equation}\label{32a}
\rho_0=\lim_{r\longrightarrow r_0}\rho(r)=\frac{1}{(1+2\lambda)D}.
\end{equation}
Thus $D=\frac{1}{(1+2\lambda)\rho_0}$ where $\rho_0$ is the energy density at the throat. We can show that
  \begin{equation}\label{32b}
H(r)=(1+\frac{D}{r^n})\rho
\end{equation}
and
  \begin{equation}\label{32c}
H_1(r)=\frac{(1-\frac{D}{r^n})}{2}\rho
\end{equation}
It is evident that the conditions $H>0$ and $H_1>0$ hold for $D>0$. Additionally, $H_2>0$ and $H_3>0$ are satisfied for $1\geq D>0$. These results demonstrate that all the ECs are satisfied for this solution. For $n=1$ and $D=1$, it is clear that
\begin{equation}\label{a33}
b(r)= \exp(1-r)
\end{equation}
which has been presented in \cite{Moa2}.  Here, we can conclude that the special case $\omega_\infty=0$ may lead to solutions with non-exotic matter. Notably, solutions with $p=r^n \rho(r)$ and $n>0$ are presented in \cite{Moa}. This class of solutions does not respect the ECs. We have shown that the case $n<0$ leads to non-exotic wormhole solutions.

The next candidate is $g(r)=\frac{D}{\ln(r)}$ which yields
\begin{equation}\label{33}
b(r)= \exp(\frac{\ln(r)^2}{2D})
\end{equation}
This shape function possesses the necessary criteria to construct a traversable wormhole. The energy density for this shape function is as follows
\begin{equation}\label{34}
\rho(r)=-\frac{\ln(r) \exp(-\frac{\ln(r)^2}{2D})}{r^3(1+2\lambda)D},
\end{equation}
which is positive for $D>0$. The function
   \begin{equation}\label{35}
H(r)=(1+\frac{D}{\ln(r)})\rho
\end{equation}
is also positive in the entire range of $r\geq r_0$ but
  \begin{equation}\label{36}
H_1(r)=(\frac{1-\frac{D}{\ln(r)}}{2})\rho
\end{equation}
is negative near the wormhole throat.  These results illustrate that the ECs are violated in this scenario.

The final choice for $g(r)$ is (\ref{8c1}) which leads to
\begin{equation}\label{37}
b(r)= r^{-\frac{2A+\ln(r)}{2D})}.
\end{equation}
Equation (\ref{4}) indicates that
 \begin{equation}\label{38}
-\frac{A}{D}<1.
\end{equation}
The form of energy density is as follows
\begin{equation}\label{39}
\rho(r)=-\frac{(\ln(r)+A)r^{-\frac{6D+\ln(r)+2A}{2D}} }{r^3(1+2\lambda)D},
\end{equation}
It is evident that for
\begin{equation}\label{38a}
\frac{A+\ln(r)}{D}>0,
\end{equation}
energy density is positive throughout the entire range of  $r\geq 1$. Conditions (\ref{38})) and (\ref{38a}) are valid for $A>0$ and $D>0$.
We can find
\begin{equation}\label{41}
\rho_0=\lim_{r\longrightarrow r_0}\rho(r)=\frac{A}{D}\frac{1}{1+2\lambda}.
\end{equation}
One can see that EoS parameter at the throat is
\begin{equation}\label{42}
\omega_0=\frac{p_0}{\rho_0}=\frac{D}{A},
\end{equation}
which must be consistent with Eq.(\ref{38}). Equations (\ref{29a1}) and (\ref{38a}) explain that NEC is satisfied when
\begin{equation}\label{43}
0<\frac{D}{A}<1
\end{equation}
is assumed. The condition (\ref{43}) is fulfilled for $A>D>0$. It is straightforward to show that other ECs are valid for this case.

We have identified several solutions for different models of $g(r)$. It was shown that certain  choices for $g(r)$ cannot yield non-exotic wormhole solutions, while others can indeed present non-exotic wormhole solutions. The relationships between the physical terms at the boundary and the parameters in the $g(r)$ functions were clarified. Let us briefly compare the solutions corresponding to different $g(r)$ functions. The key point is that the value of $\omega_\infty$ plays a crucial role in the form of shape function. Additionally, it can be concluded that the free parameters $A$ in (\ref{8c1}) address the singularity issue in the EoS parameter at the wormhole throat. The parameter $D$ helps us in finding solutions with non-exotic matter. It is straightforward to see that the $n$ in the (\ref{b10})  affects the rate of convergence to a linear EoS.

\section{Concluding remarks}

Given that the equation of state of the wormhole is not well understood and constrained, we can attempt to use a more general EoS instead of a linear EoS . In this work, the EoS is asymptotically linear which is of great interest. As it was mentioned in \cite{variable}, variable EoS may appear more physical as the linear EoS stands as a global equation; however, at a local level, there is no need to adhere strictly to a linear equation. This concept can be linked to the unique geometry of a wormhole in close proximity to its throat. The variable EoS parameter has injected fresh vitality into the exploration of wormhole physics when compared to a constant parameter. It reduces the exotic matter in the context of GR and also helps us to present a large class of non-exotic solutions in the context of $f(R,T)$.

Although, experimental detection of wormholes has not yet been done, in this study, our investigation has focused on the possibility of wormhole geometry in the context of $f(R, T)$ gravity. In wormhole theory, the violation of ECs poses a fundamental inconsistency that needs to be dealt with.  In the context of $f(R,T)$, exotic matter is just replaced by an equivalent modified form of gravity. Many forms for $f(R,T)$ function  have been considered in the literature. The wormhole in the $f(R,T)$ theory with linear function $f(R,T)=R+2\lambda T$ has been extensively studied by many researchers. Most of these solutions violate the ECs when a linear function, $f(R,T)=R+2\lambda T$, is considered.  In this context, wormholes with a linear EoS form a category of solutions that adhere to the ECS \cite{Azizi}. The satisfaction of ECs relies on the factor $1+2\lambda$. This parameter can eliminate the influence of exotic matter in the field equations. In standard GR, $2\lambda T$ is absent, so the flaring-out condition results in the exotic matter. With the presence of $2\lambda T$ the Einstein gravitational constant shifts to $1+2\lambda$. The negative sign for this effective gravitational constant could address the issue of exotic matter. To summarize, within the framework of GR, solutions with negative energy density, which violate the lateral and radial NEC components, can fulfill the ECs in the context of $f(R,T)$. We have shown that the negative sign of $1+2\lambda$ leads to negative pressure at the wormhole throat in the context of $f(R,T)$.

According to our knowledge, there are limited solutions in the existing literature that satisfy the ECs for $f(R,T)=R+2\lambda T$. We analyzed various prominent shape functions for (\ref{14}). However, none of these shape functions successfully demonstrated a non-exotic wormhole solution except for $b(r)=\frac{a^r}{a}$. In this work, we explore a large class of solutions in the context of $f(R,T)$ gravity while $f(R,T)=R+2\lambda T$ and $L_m=-\rho$ are assumed.
We have developed a comprehensive framework to uncover exact solutions, surpassing the previous model by focusing on  choosing a variable EoS. We use an asymptotically linear EoS for the wormhole fluid. To enhance the viability of our solutions, we suggest a positive energy density.

It was shown that the model parameters' numerical values play a major role in determining the results. We have shown that the free parameter in the shape function  can be related to the values of the energy-momentum tensor at the throat or boundary of the wormhole. In other words,  the boundary is sensitive to changes in parameters
of the shape function and EoS parameter.

 Relevant $g(r)$ functions have been explored, with some resulting in viable wormhole solutions.
We explored two classes of asymptotically flat solutions. In the first class, we examined solutions with $\omega_\infty\neq 0$. For the second class of solutions, $\omega_\infty=0$ was analyzed. It was demonstrated that these two classes of solutions may lead to different shape functions for the same $g(r)$, which vary in the ECs status. We can see that the rate of coherence to a linear EoS depends on the form of $g(r)$. For example, $g(r)=\frac{D}{r^n}$ for $n>1$ tends faster to a liner EoS in contrast to $g(r)=\frac{D}{r}$. Also, it was shown that the EoS parameter at large distance ($\omega_\infty$) depends on the free parameters in the shape function.

The theoretical coherence and rationale behind these extensions of $f(R,T)$ can be elucidated to delve into new possibilities in wormhole theory and cosmological forecasts of $f(R,T)$ theory. Along this way, we have considered a vanishing  redshift function, i.e., $\phi(r) = 0$, but solutions with non-constant redshift function can be explored. Additionally, our algorithms can also be applied to various models of $f(R,T)$ or alternative shape function formats.

\end{document}